%% file: diode.tex
\documentclass[ 
reprint,
amsmath,
amssymb,
groupedaddress,
aps,
floatfix,
nofootinbib,
twocolumn,
prb,
superscriptaddress,
a4paper
]{revtex4-2}
\usepackage{placeins}  
\usepackage[normalem]{ulem}
\usepackage[dvipsnames]{xcolor}
\usepackage{graphicx}
\usepackage{dcolumn}
\usepackage{bm}
\usepackage{svg}
\usepackage{url}
\usepackage{soul}
\usepackage{orcidlink}
\usepackage{changes}
\usepackage{hyperref}

\begin{document}


\title{Diode Effect in Nonlinear High-Kinetic-Inductance Transmission Line Resonators}

\author{Niklas Gaiser\,\orcidlink{0000-0002-8539-0292}}
\email{niklas.gaiser@uni-ulm.de}
\author{Ciprian Padurariu\,\orcidlink{0000-0001-9568-2080}}
\affiliation{
Institute for Complex Quantum Systems and IQST, University of Ulm, 89081 Ulm, Germany
}

\author{Bj\"orn Kubala\,\orcidlink{0000-0001-6685-0233}}
\affiliation{
Institute for Complex Quantum Systems and IQST, University of Ulm, 89081 Ulm, Germany
}
\affiliation{
German Aerospace Center (DLR), Institute of Quantum Technologies, 89081 Ulm, Germany
}

\author{Nadav Katz\,\orcidlink{0000-0002-9264-078X}}
\affiliation{
Racah Institute of Physics, Hebrew University of Jerusalem, Jerusalem 9190401, Israel
}
\affiliation{
Qarakal Quantum Ltd., Derech Menachen Begin 144, Tel Aviv 6492102, Israel
}

\author{Joachim Ankerhold\, \orcidlink{0000-0002-6510-659X}}
\affiliation{
Institute for Complex Quantum Systems and IQST, University of Ulm, 89081 Ulm, Germany
}

\date{\today}

\input{sections/abstract}
\maketitle

\input{sections/introduction}

\input{sections/theory}

\input{sections/results}

\input{sections/conclusion}
~\\
\input{sections/acknowledgments}
\appendix
\input{sections/appendix_a}


\bibliography{bib.bib}

\end{document}

%% file: sections/abstract.tex
\begin{abstract}

High-kinetic-inductance (HKI) transmission lines provide a promising platform for compact nonlinear superconducting microwave devices. Existing theoretical descriptions are typically based on the slowly varying envelope approximation, which becomes inadequate for strongly nonlinear resonant structures with pronounced spatial field variations. Here, we develop a theoretical framework for single-frequency nonlinear wave propagation in HKI transmission lines by formulating the problem as a boundary value problem that retains the full spatial dependence of the electromagnetic fields. Applying the method to resonant transmission-line geometries, we demonstrate power-dependent resonance shifts, bistable transmission solutions, and strongly direction-dependent transport arising from asymmetric impedance barriers, yielding transmission contrasts of up to $93\%$ without magnetic bias fields. The framework is further extended to a three-port stub geometry, where nonlinear interference produces shifted anti-resonances and Duffing-like spectral distortions. Our approach provides a versatile tool for the analysis and design of strongly nonlinear superconducting microwave devices.

\end{abstract}

%% file: sections/introduction.tex
\section{Introduction\label{sec:introduction}}

Superconducting circuits provide a versatile platform for a wide range of quantum technologies, including quantum computation, quantum sensing, and quantum-limited microwave measurement and amplification \cite{devoret2013,blais2020,gu2017}. Their appeal stems from the ability to realize highly controllable microwave systems that combine low dissipation with strong nonlinear response \cite{devoret2013,blais2020}.

Nonlinear microwave devices play a central role in applications such as parametric amplification and frequency conversion for quantum-limited signal processing
\cite{yurke1989,lehnert2007,macklin2015}. 
Consequently, the development of nonlinear circuit elements combining large dynamic range, low dissipation, and broadband operation is of considerable interest for superconducting microwave technologies. Such nonlinearities can arise either from intentionally engineered circuit elements, such as Josephson junctions (JJs), or from the intrinsic nonlinear kinetic inductance of the superconducting material itself \cite{makhlin2001,annunziata2010}. In an appropriate power regime, transmission lines (TLs) fabricated from high-kinetic-inductance (HKI) materials exhibit pronounced nonlinear behavior \cite{eom2012,vissers2016}. Their potential for nonlinear microwave applications has recently been demonstrated in high-kinetic-inductance resonators exhibiting strong Kerr nonlinearities \cite{gao2008,frasca2023,yang2024}. Compared to Josephson-junction-based nonlinearities, HKI transmission lines provide a distributed nonlinear medium with broadband operation, high power handling capability, and comparatively simple fabrication \cite{eom2012, vissers2016, gao2017, malnou2021, goldstein2020, goldstein2022}. 
Their compact footprint further makes them attractive for the realization of scalable on-chip microwave devices.

Previous theoretical descriptions of nonlinear HKI transmission lines have often relied on perturbative approaches based on the slowly varying envelope approximation \cite{erickson2017, malnou2021, kern2023}. While this framework is well suited for weak nonlinearities and nearly uniform wave propagation, its accuracy can deteriorate in the strongly nonlinear regime. In particular, resonant structures and devices containing impedance discontinuities can exhibit rapid spatial variations of the field amplitude and substantial nonlinear phase accumulation, thereby violating the assumptions underlying the perturbative treatment. Unlike conventional coupled-mode descriptions, the approach presented here retains the full spatial dependence of the electromagnetic fields and is therefore applicable to strongly resonant structures with large field gradients and impedance discontinuities.

Beyond their fundamental interest, nonlinear transmission lines offer a promising route toward direction-dependent microwave transport. Nonreciprocal microwave components are essential for protecting sensitive quantum devices from environmental noise and unwanted back-propagating signals. Conventional nonreciprocal elements often rely on magnetic materials, external magnetic bias fields, or optomechanical interactions \cite{bernier2017,abdo2017,lecocq2017}, which are difficult to integrate into larger networks of superconducting quantum circuits. Nonlinear asymmetric structures therefore provide an attractive route toward direction-dependent transmission without the need for magnetic biasing \cite{cotrufo2021}.

In this work, we develop a self-consistent framework for treating strongly nonlinear HKI transmission lines beyond the slowly varying envelope approximation. Using this approach, we investigate several transmission-line geometries containing nonlinear resonant sections and impedance discontinuities. We demonstrate that asymmetric resonator barriers lead to direction-dependent field build-up within the resonator, resulting in unequal nonlinear phase shifts and resonance shifts for opposite propagation directions. Furthermore, we observe bistable transmission solutions and pronounced transmission asymmetries that emerge from the interplay of nonlinear self-interaction, resonant field enhancement, and structural asymmetry. As a consequence, the transmission characteristics become strongly direction dependent, giving rise to effective isolation and asymmetric power transmission.

The presented results provide a versatile framework for investigating strong nonlinear wave propagation in superconducting transmission structures and demonstrate how distributed kinetic-inductance nonlinearities can be harnessed to engineer direction-dependent microwave transport.

%% file: sections/theory.tex
\section{Theory\label{sec:theory}}
\subsection{Nonlinear Transmission 
Line}
\begin{figure}[htbp]
    \centering
    \includegraphics[width=1\linewidth]{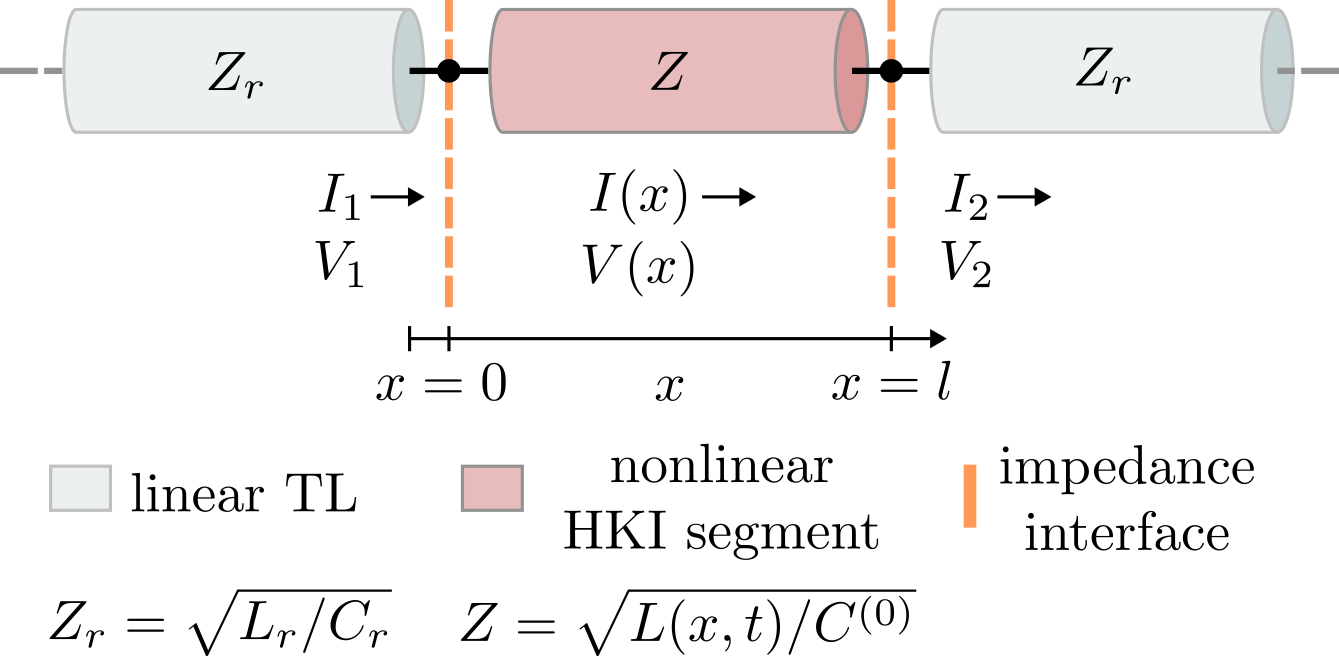}
    \caption{Schematic layout of nonlinear transmission line resonator: Within a transmission line with impedance $Z_r$ a section is replaced by a segment of length $l$ with nonlinear impedance $Z$ as in Eq.~\eqref{eq:impedance}. The voltages $V_1$, $V_2$ and $V(x)$ as well as the currents $I_1$, $I_2$ and $I(x)$ have to fulfill the Kirchhoff rules at the interfaces ($x=0,l)$.}
    \label{fig:setup1}
\end{figure}

The fundamental equations of a transmission line are the telegrapher's equations \cite{pozar,oates1993}. In the lossless case and with constant capacitance per unit length $C^{(0)}$, they are given by
\begin{subequations}
\begin{eqnarray}
\frac{\partial }{\partial x}V(x,t)&=&- \frac{\partial}{\partial t} \big[ L(x,t) I(x,t) \big] \label{eq1a}
\\
\frac{\partial }{\partial x}I(x,t)&=&-  C^{(0)} \frac{\partial }{\partial t} V(x,t) \label{eq1b}.
\end{eqnarray}\label{eq1}
\end{subequations}
Transmission lines realized by high-kinetic-inductance superconducting micro-strips, such as in Ref.\,\cite{eom2012,vissers2016,gao2017,goldstein2020,goldstein2022}, have a nonlinear inductance per unit length $L(x,t)$ that arises due to its kinetic contribution
\begin{equation}
    L(x,t) = L^{(0)} \left[1 + \frac{I^2(x,t)}{I_\ast^2} \right],\label{eq:kineticinductance}
\end{equation}
with negligible geometrical inductance.
The parameter $I_\ast$ describes the scale of the nonlinearity and is of the order of the critical current of the superconducting strip-line \cite{zmuidzinas2012}. This position-time-dependent inductance determines the nonlinear impedance
\begin{equation}
\label{eq:impedance}
    Z=Z^{(0)} \sqrt{\frac{L(x, t)}{L^{(0)}}}\ \ , \ \ Z^{(0)} = \sqrt{\frac{L^{(0)}}{C^{(0)}}}\, 
\end{equation}
as seen in Fig.~\ref{fig:setup1}.

Following the approach in \cite{oates1993}, we employ a single frequency (harmonic balance) ansatz for voltage and current waves.
This approximation assumes that the nonlinear response is dominated by the fundamental probe frequency, while higher harmonics and frequency mixing generated by the nonlinear inductance remain sufficiently weak to be neglected. Under these assumptions, the position and time dependencies can be separated according to
\begin{subequations}
\begin{eqnarray}
V(x,t) &=&   V_s(x) \sin \left( \omega t \right) + V_c(x) \cos \left( \omega t \right) \label{eq4a}\\
I(x,t) &=&  I_s(x) \sin \left( \omega t \right) + I_c(x) \cos \left( \omega t \right). \label{eq4b}
\end{eqnarray}\label{eq4}
\end{subequations}
The resulting nonlinear equations are
\begin{subequations}
\begin{eqnarray}
    \frac{\partial V_s}{\partial x} &=& \omega L^{(0)} \left[ I_c + \xi I_c \left(  I_s^2  + I_c^2 \right) \right] \label{eq5a}\\
    \frac{\partial V_c}{\partial x} &=& - \omega L^{(0)}  \left[ I_s + \xi I_s \left(  I_s^2  + I_c^2 \right) \right] \label{eq5b}\\
    \frac{\partial I_s}{\partial x} &=&   \omega C^{(0)} V_c \label{eq5c}\\
    \frac{\partial I_c}{\partial x} &=& - \omega C^{(0)}  V_s   \label{eq5d},
\end{eqnarray} \label{eq:5}
\end{subequations}
where it is convenient to define the nonlinearity parameter $\xi= 3/(4 I_\ast^2)$. The approximation made describes the nonlinear response of the transmission line only to first order in $\xi\left(  I_s^2  + I_c^2 \right)$ and therefore is valid only for signals with moderate power $P \ll Z I_\ast^2$. Nevertheless, even to the leading order considered here, the effect of the nonlinearity is non-perturbative, as it may accumulate along the full length of the transmission line, yielding large deviations from a transmission line with linear inductance.

The resulting simplified equations form a set of time-independent ordinary differential equations that can be summarized as two coupled second-order differential equations, 
\begin{subequations}
\begin{eqnarray}
    \frac{\partial^2 I_s}{\partial x^2} &=&  - \beta^2  \left[ I_s + \xi I_s \left(  I_s^2  + I_c^2 \right) \right] \label{eq6a}\\
    \frac{\partial^2 I_c}{\partial x^2} &=& - \beta^2    \left[ I_c + \xi I_c \left(  I_s^2  + I_c^2 \right) \right], \label{eq6b}
\end{eqnarray} \label{eq6}
\end{subequations}
where 
\begin{equation}
    \beta(\omega)=\omega\sqrt{L^{(0)}C^{(0)}} \label{eq7}
\end{equation}
is the propagation constant of the linear transmission line. 

\begin{figure}[htbp]
\centering
\includegraphics[width=1\linewidth]{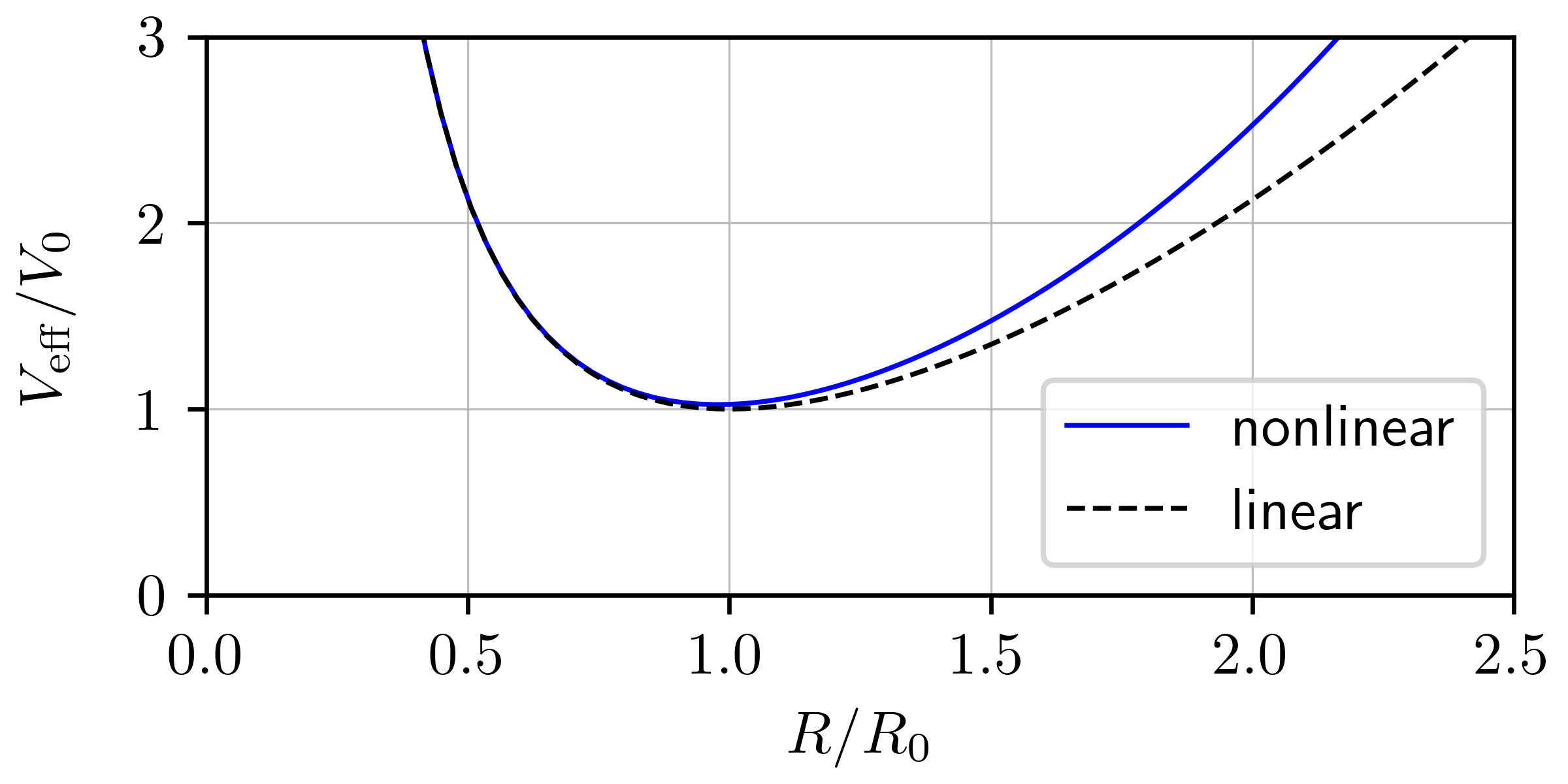}
\caption{Effective one-dimensional potential for the linear case $\xi = 0$ (dashed black) and the nonlinear case $\xi > 0$ (solid blue).}
\label{fig:D0}
\end{figure}

Theses equations can be mapped to a mechanical analogue, where we interpret the spatial derivatives as time derivatives instead. In this picture, we write $\dot{I}_{s,c}$ to express $\partial I_{s,c}/\partial x$ and consider $\{\dot{I}_{s, c}, I_{s, c}\}$ as pairs of variables to formulate the Lagrange function $\mathcal{L}(I_{s,c}, \dot{I}_{s,c})$ that describes the motion of a fictitious particle with unit mass and coordinates $(I_s,I_c)$ moving in the rotational-invariant potential
\begin{equation}
    V_\mathrm{pot}(I_s,I_c) =  \frac{\beta^2}{2  } \left[ \left( I_s^2 + I_c^2 \right) + \frac{\xi}{2} \left(  I_s^2  + I_c^2\right)^2 \right]. \label{eq8}
\end{equation}
To proceed, it is convenient to transform to polar coordinates defined by \hbox{$R= \sqrt{I_s^2 + I_c^2}$} and $\theta= \arctan ( I_s / I_c )$. This yields the Lagrangian
\begin{equation}
    \mathcal{L}(R, \dot{R}, \dot{\theta}) = \frac{1}{2} \dot{R}^2  + \frac{1}{2} R^2  \dot{\theta}^2  -  \frac{\beta^2}{2 } \left( R^2 + \frac{\xi}{2} R^4 \right)\label{eq:langangian}
\end{equation}
which apparently does not carry an explicit $\theta$-dependence. 
Accordingly, we find a constant of motion that we associate with the conserved angular momentum
\begin{equation}
    L_\theta =  R^2 \dot{\theta}. \label{eq10}
\end{equation}
Returning to the current and voltage representation, this constant $L_\theta = \omega C^{(0)} \mathrm{Re} \left[ I(x) V^\ast(x) \right]$ is in fact proportional to the time-averaged power flow at position $x$ \cite{pozar}, thus representing  the current conservation (continuity equation). Further, from Eq.~\eqref{eq:langangian} we can extract an effective one-dimensional potential of the form
\begin{equation}
        V_{\mathrm{eff}}(R) = \frac{1}{2} \frac{L_\theta^2}{R^2} + \frac{\beta^2}{2}\left(R^2 + \frac{\xi}{2} R^4 \right). \label{eq11}
\end{equation}
Note that the nonlinear term steepens the effective potential at larger amplitudes, introducing a contribution that scales as $R^4$, as illustrated in Fig.~\ref{fig:D0}.\\
Using this transformation, the two coupled equations in Eq.~\eqref{eq6} can be reduced to a single nonlinear equation of motion,
\begin{equation}
\frac{\partial^2 R}{\partial x^2} = \frac{L_\theta^2}{R^3} - \beta^2 \left( R + \xi R^3 \right).
\label{eq12}
\end{equation}
This `dynamics' of a fictitious particle determines completely the field propagation inside the nonlinear section of the setting illustrated in Fig.~\ref{fig:setup1}. An essential ingredient are the boundary conditions that we will discuss in the next section.

\subsection{Boundary Value Problem}
If the voltage and current are specified at a given position along the transmission line, Eq.~\eqref{eq12} can be integrated directly to obtain the fields at any other position. In a scattering experiment, however, the situation is generally different.

\begin{figure}[htbp]
\centering
\includegraphics[width=1\linewidth]{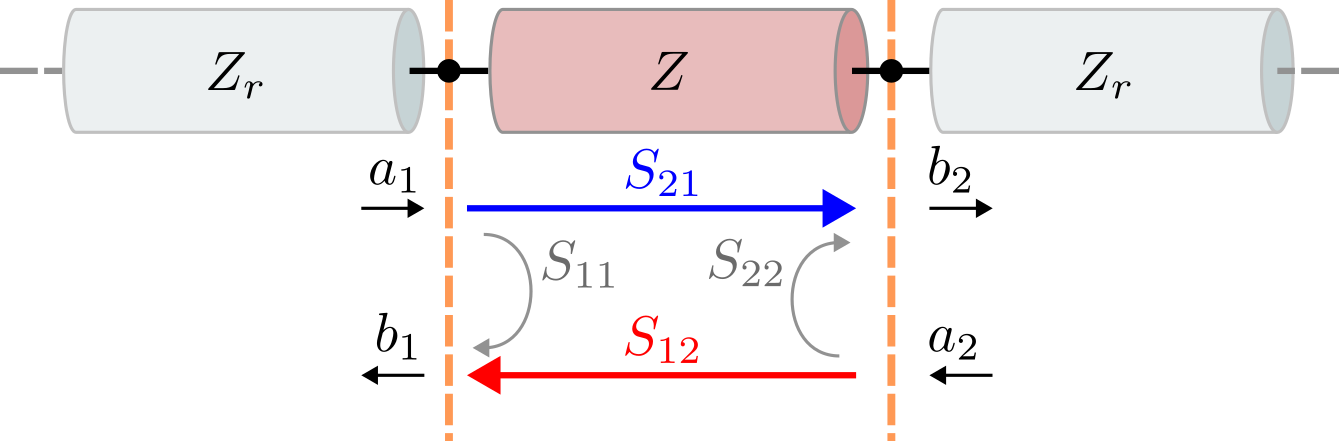}
\caption{Two-port scattering problem for a nonlinear resonator. Incident signals $a_1$ and $a_2$ enter the device through the left and right ports, respectively. The corresponding scattered outputs are denoted by $b_1$ and $b_2$. The scattering parameters $S_{21}$ and $S_{12}$ describe the transmission through the device from port 1 to 2 and from port 2 to 1 respectively. The reflection at each port is described by the parameters $S_{11}$ and $S_{22}$.}
\label{fig:D1}
\end{figure}

A typical two-port scattering configuration is illustrated in Fig.~\ref{fig:D1}. Prior to the scattering process, only the incident signals $a_1$ and $a_2$ at the two device ports are known, whereas the outgoing signals $b_1$ and $b_2$ are determined by the scattering process itself and therefore constitute part of the solution. Since the conditions are specified at both ends of the device rather than at a single point, the problem must be formulated as a boundary value problem rather than an initial value problem. Consequently, a different solution strategy is required.

The voltage and current at the device ports are related to the incident and scattered wave amplitudes through $V_n = a_n + b_n$ and $Z_r I_n = a_n - b_n$, where the index $n$ denotes the corresponding port.

For an impedance discontinuity between transmission lines with characteristic impedances $Z_r$ and $Z$, respectively, the reflection and transmission coefficients at a single interface are given by $\tilde{S}_{11}=(Z - Z_r)/(Z_r + Z)$ and $\tilde{S}_{21}=2 \sqrt{Z_r Z}/(Z_r + Z)$. To obtain the scattering matrix across the entire structure including one or several nonlinear parts (with impedances $Z_j$, $j=$1, 2, 3, \ldots) terminated by regular parts (with impedance $Z_r$), one has to solve Eq.~\eqref{eq12} and infer for given input $a_1, a_2$ from $R, \theta$ the elements $S_{21}=b_2/a_1$, $S_{12}=b_1/a_2$ and $S_{11}=b_1/a_1$. 
The strength of the impedance mismatch determines the magnitude of the reflection coefficient and therefore the sharpness of the resonances formed within the device. For a reciprocal linear system, the transmission coefficients satisfy $S_{12}=S_{21}$. In order to obtain direction-dependent transmission, the combined scattering response of the nonlinear structure must differ for opposite propagation directions, resulting in $S_{12}\neq S_{21}$.

Taking into account the impedance mismatch at both ports, the boundary value problem can be solved using a shooting method \cite{keller2018,ascher1995}. For a given complex input amplitude $a_1$, an initial guess is made for the complex output amplitude $b_1$. The corresponding field is then propagated through the device to the opposite port. The value of $b_1$ is iteratively adjusted until the propagated solution satisfies the prescribed boundary condition $a_2$.

In the common case $a_2=0$, it is more convenient to instead guess the output amplitude $b_2$ and propagate the solution backward through the device. Owing to the global phase symmetry of Eq.~\eqref{eq12}, the transformation
$\theta(x) \rightarrow \theta(x)+  \theta_0$ leaves the nonlinear dynamics invariant. Consequently, the overall phase of the solution can be chosen arbitrarily, and only the magnitude $|b_2|$ must be determined by the shooting procedure. Once a self-consistent solution has been obtained, the global phase can be adjusted such that the resulting input signal acquires the desired phase of $a_1$. The boundary value problem therefore reduces to a one-dimensional root-finding problem.

The same procedure can be extended to devices containing multiple nonlinear segments by starting from the rightmost output and propagating the solution backward through the structure segment by segment. As discussed later, a single real-valued shooting parameter is likewise sufficient for the stub geometry. In contrast, coupled multi-line devices such as those investigated in \cite{goldstein2022}, or more generally structures containing closed loops, require multiple independent shooting parameters and therefore pose a substantially greater numerical challenge.

\subsection{Diode effect} 
\begin{figure}[htbp]
    \centering
    \includegraphics[width=1\linewidth]{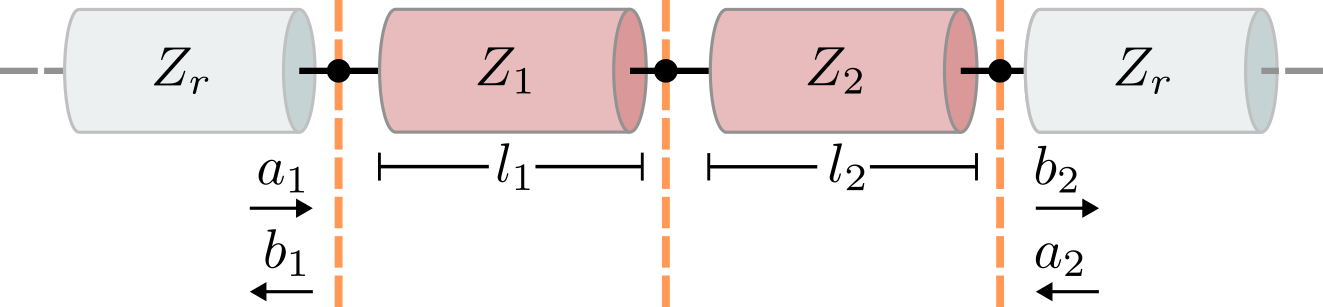}
    \caption{Schematic layout nonreciprocal device: In a transmission line with $Z_r$ a section is replaced by two segments with $Z_1$ and $Z_2$ ($Z_1 \neq Z_2$), which have nonlinear kinetic inductances. At their mismatch the in-/output of the segment with $Z_1$ translates to the out-/input of the segment of $Z_2$.}
    \label{fig:D2}
\end{figure}

Now that we know how to address a nonlinear segment within a transmission line, which has potentially mismatched impedance, we want to employ this to propose a device that shows nonreciprocal behavior, see Fig.~\ref{fig:D2}.

In fact, as a result of the nonlinear section's self-interaction, it is enough to break the left-right symmetry in a device in order to achieve this. To realize different reflection coefficients at the opposite device ports with identical impedance to ground, one could adiabatically change the impedance of the central segments. Indeed, it will turn out to be sufficient (and more convenient) to split the central section into two parts separated by a slight impedance mismatch that introduces minimal reflection only.
In the linear case, the order in which the signal propagates through the interfaces is irrelevant. In the nonlinear case, the order matters, since the signal in the nonlinear section interacts with itself. Therefore, a signal first passing through an interface with a small impedance difference and then through an interface with a high impedance difference will show a large influence of the nonlinearity. On the other hand, a signal first passing a high impedance difference and then a low one will have a low influence of the nonlinearity. This breaks the left-right symmetry of the device and causes a nonreciprocal effect. This effect is more prominent the more the reflection coefficients at the interfaces differ from each other and the higher the signal input power and thus the nonlinear contribution.

\begin{figure}[htbp]
    \centering
    \includegraphics[width=1\linewidth]{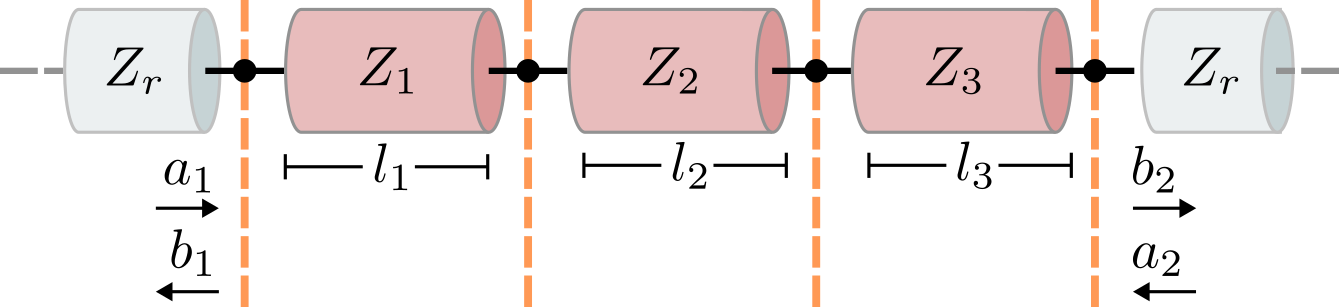}
    \caption{Schematic layout of nonreciprocal device with three unmatched impedance sections: Three segments are inserted with unequal impedances $Z_1$, $Z_2$ and $Z_3$ and nonlinear features.}
    \label{fig:D3}
\end{figure}

It is possible to extend nonreciprocal devices, such as presented above, by more segments, see Fig.~\ref{fig:D3}. This changes the transmission spectrum and creates more resonances in the device. Thus, it is possible to design desired features for special purposes such as diodes.

%% file: sections/results.tex
\section{Results}\label{sec:results}
First, we numerically simulate the transmission spectra of the devices introduced above. To investigate the effect of left-right symmetry breaking, we apply an input signal with amplitude $|a_1| = 1$ on one side of the device while setting the opposite input to $|a_2| = 0$. We then reverse these boundary conditions and compare the resulting transmission characteristics.

\begin{figure}[htbp]
    \centering
    \includegraphics[width=1\linewidth]{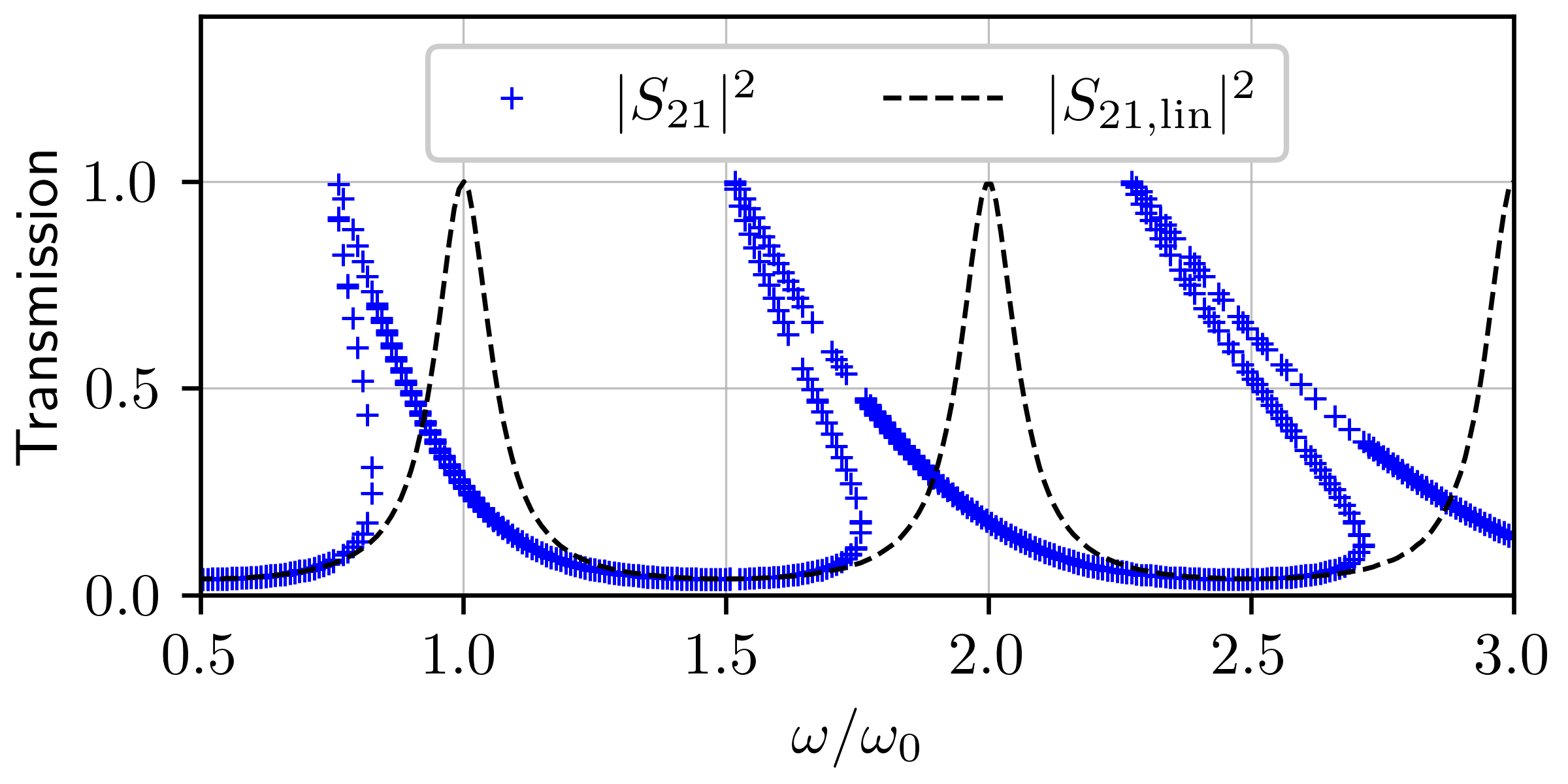}
    \caption{Simulated transmission spectrum of setting in Fig.~\ref{fig:D1}. The resonator section has an impedance of $Z^{(0)}/Z_r = 10 $ and  its length is chosen such that $\beta(\omega_0)l=\pi$, where $\omega_0$ is a reference frequency set to the linear device's resonance. In the linear case (dashed black), a sequence of Fabry-Pérot resonances is observed. In the nonlinear case (blue), the resonances shift toward lower frequencies. The nonlinearity parameter is $\xi = 3/(4I_\ast^2) = 0.01$, where $I_\ast$ is expressed in units of the input current $a_1/Z_r$. Higher-frequency resonances exhibit larger shifts than lower-frequency resonances. In certain frequency ranges, up to three distinct transmission solutions exist, one of which can be expected to be unstable.}
    \label{fig:R1}
\end{figure}

The resonator shown in Fig.~\ref{fig:D1} exhibits Fabry-Pérot-like resonances, as shown in Fig.~\ref{fig:R1}. In the nonlinear regime, the resonance frequencies shift toward lower values, and the magnitude of the shift increases with the resonance order. Depending on the strength of the nonlinearity, the shift can become sufficiently large that multiple solutions exist for a single frequency. Although all these solutions are formally valid within the single-frequency approximation of Eq.~\eqref{eq6} applied to the original partial differential equations Eq.~\eqref{eq1}, one generally expects additional solutions to emerge in pairs consisting of one stable and one unstable branch, similar to the behavior of a Duffing oscillator \cite{nayfeh2024,guckenheimer1983}. Further, hysteretic behavior can be expected when parameters such as the frequency are swept through the multivalued regime. For the symmetric device, the transmission response is identical for both propagation directions.

\begin{figure}[htbp]
\centering
\includegraphics[width=1\linewidth]{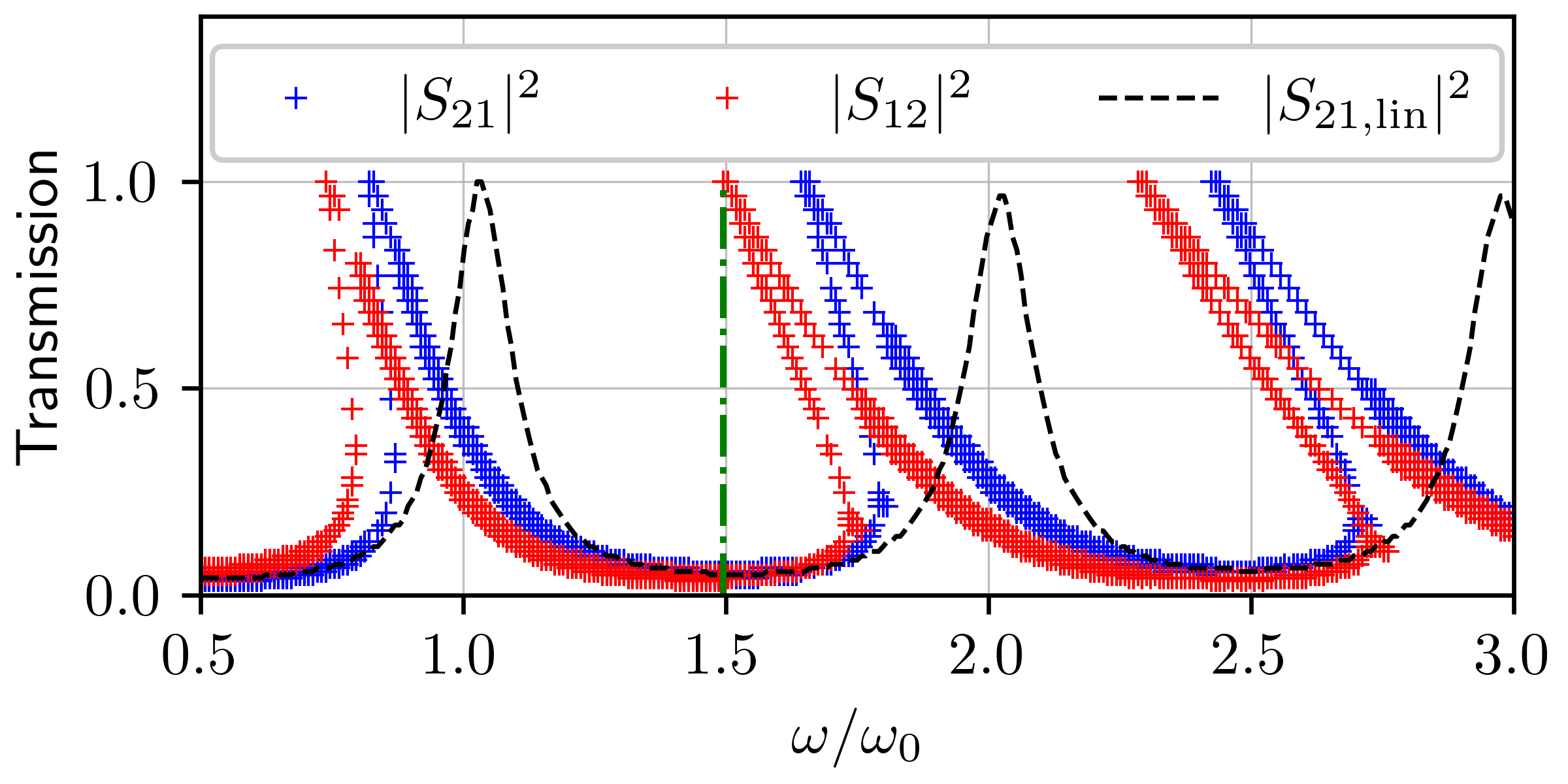}
\caption{Simulated transmission spectrum of the asymmetric device shown in Fig.~\ref{fig:D2}. The resonator section consists of two segments with impedances $Z_1^{(0)}/Z_r = 10$ and $Z_2^{(0)}/Z_r = 8$. Their lengths are chosen such that $\beta_1 (\omega_0) l_1 = 4\pi/5$ and $\beta_2(\omega_0) l_2 = \pi/5$, and both segments possess a nonlinearity of $\xi_{1,2}=0.01$. The transmission spectra for propagation from left to right, $S_{21}$ (blue), and from right to left, $S_{12}$ (red), exhibit pronounced differences due to the direction-dependent nonlinear resonance shifts. For comparison, the linear transmission spectrum $S_{21,\mathrm{lin}}$ (dashed black) is identical for both propagation directions. At $\omega/\omega_0 = 1.5$ (dashed-dotted green) the contrast reaches $\eta = 93\%$. 
}
\label{fig:R2}
\end{figure}

In contrast, a device containing two unequal resonator barriers exhibits direction-dependent field build-up. As argued above, we realize such barriers by introducing two  central segments, with a small impedance mismatch, see Fig.~\ref{fig:D2}, where  transmission across the central mismatch is near perfect, $t_c=0.993$, while the coefficients for left ($t_l\approx 0.575 $) and right ($t_r\approx 0.629$) barrier differ significantly. The current amplitudes in the two central segments thus depend on whether the propagating signal encounters the larger  or the smaller reflection coefficient first, as shown in Fig.~\ref{fig:R2}.

Consequently, the nonlinear inductance affects the two propagation directions differently, leading to unequal transmission characteristics. Indeed, the resonance shift is more pronounced when the signal first encounters the lower-reflection barrier (from $Z_r$ to $Z_2$) and subsequently the higher-reflection barrier (from $Z_1$ to $Z_r$). For propagation in the opposite direction, the shift is reduced.
To quantify the transmission asymmetry at a given operating point, we define the transmission contrast as
\begin{equation}
\eta = \max \left| \frac{|S_{21}|^2-| S_{12}|^2}{|S_{21}|^2+|S_{12}|^2}\right|.
\end{equation}
For the parameters considered in Fig.~\ref{fig:R2}, the device exhibits a transmission contrast of approximately $\eta = 93\%$ at $\omega/\omega_0 = 1.5$.
We note in passing, that in general, a resonator composed of two segments with different lengths supports several distinct fundamental resonances, which give rise to transmission peaks of varying heights. For the parameters considered in Fig.~\ref{fig:R2}, however, this is only weakly visible and manifests itself merely as minute modifications of the linear spectrum compared to the regular Fabry-Pérot resonance pattern observed in Fig.~\ref{fig:R1}

\begin{figure}[htbp]
\centering
\includegraphics[width=1\linewidth]{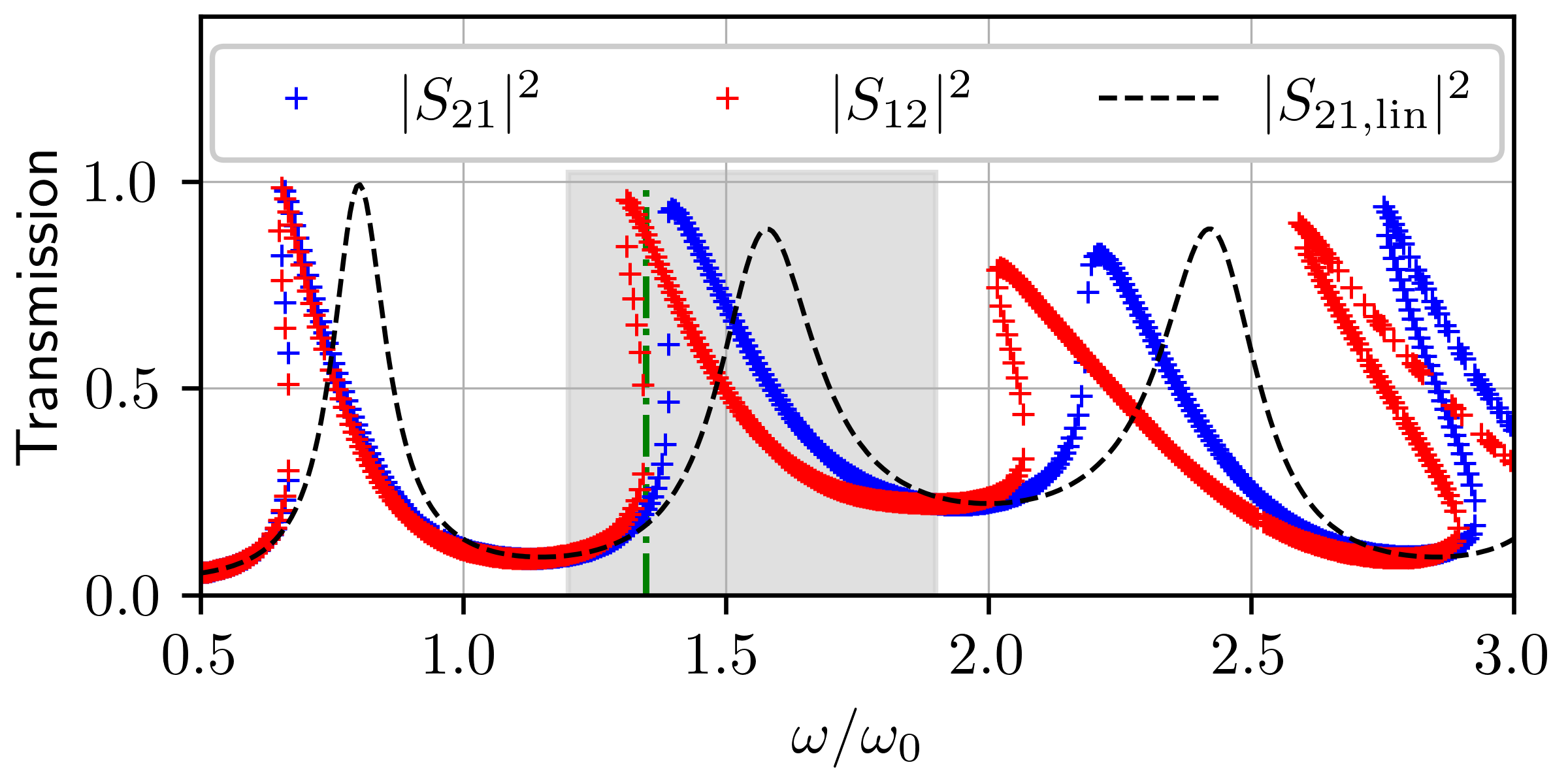}
\caption{Simulated transmission spectrum of the device shown in Fig.~\ref{fig:D3}. The three central sections are nonlinear with $\xi_n = 0.01$. Their lengths are $\beta_1(\omega_0) l_1 = \pi/4$, $\beta_2(\omega_0) l_2 = 4\pi/5$, and $\beta_3(\omega_0) l_3 = \pi/4$, with corresponding impedances $Z_1^{(0)}/Z_r = 8$, $Z_2^{(0)}/Z_r = 10$, and $Z_3^{(0)}/Z_r = 2$. In the linear case (dashed black), the device exhibits resonances arising from the different propagation path lengths. The resonance peaks display varying heights and widths due to the unequal reflection coefficients. In the nonlinear case, the resonance shifts depend on the propagation direction and are more pronounced for backward propagation (red) than for forward propagation (blue). At the marked frequency $\omega/\omega_0 = 1.35$ (dashed-dotted green) the contrast is $\eta = 61\%$. A magnified view of the gray-highlighted region is shown in Fig.~\ref{fig:R4}.}
\label{fig:R3}
\end{figure}

Having established a device exhibiting a substantial diode effect, we next consider strategies for further enhancing its performance. For the device containing three impedance-mismatched segments, shown in Fig.~\ref{fig:D3}, we obtain a considerably more intricate transmission spectrum, as illustrated in Fig.~\ref{fig:R3}. This behavior is expected, since the device supports resonances associated with multiple propagation paths arising from scattering at the impedance interfaces. Introducing an additional segment increases the number of tunable design parameters and thereby provides greater flexibility for tailoring the transmission characteristics. Consequently, further optimization of the segment lengths, impedances, and nonlinearities could enhance the transmission contrast and broaden the operational bandwidth while maintaining a comparatively simple device architecture and low fabrication complexity.

For the device shown in Fig.~\ref{fig:D3} whose transmission is shown in Fig.~\ref{fig:R3}, we obtain a transmission contrast of $\eta = 61\%$ at $\omega / \omega_0 = 1.35$, but in distinction to the example in Fig.~\ref{fig:R2} this value is taken at a frequency where only a single stable solution exists for each propagation direction, thereby avoiding bistability.

\begin{figure}[htbp]
    \centering
    \includegraphics[width=1\linewidth]{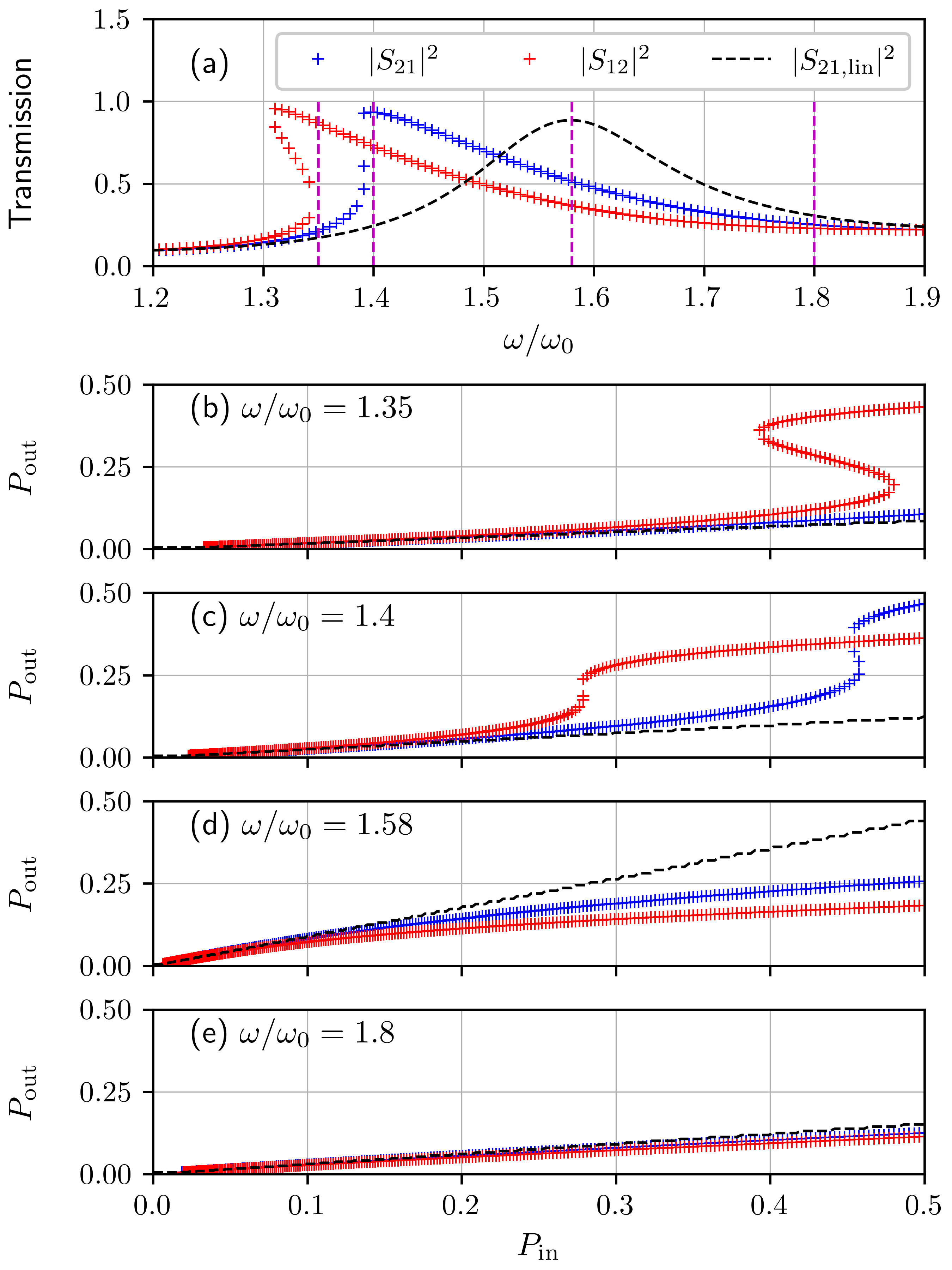}
\caption{Simulated power response of the device shown in Fig.~\ref{fig:R3}. The output power $P_{\mathrm{out}}$ is plotted as a function of the input power $P_{\mathrm{in}}$ for propagation from left to right (blue) and from right to left (red). For comparison, the corresponding linear response is shown as a dashed black line. Panel (a) displays the transmission spectrum at $P_{\mathrm{in}} = 0.5$, with the frequencies investigated in panels (b)-(e) indicated by dashed magenta lines. Panels (b)-(e) show the power response at $\omega/\omega_0 = 1.35$, $1.4$, $1.6$, and $1.8$, respectively. Depending on frequency, the nonlinear response exhibits bistability, direction-dependent threshold behavior, resonance suppression, and nearly linear transmission far from resonance.}
\label{fig:R4}
\end{figure}

The nonlinear contribution to the kinetic inductance increases with the input power $P_{\mathrm{in}}$. For the multi-segment resonator shown in Fig.~\ref{fig:R3}, we therefore examine the output power as a function of $P_{\mathrm{in}}$ for both propagation directions. The corresponding results at several representative frequencies are presented in Fig.~\ref{fig:R4}.

The selected frequencies probe distinct regions of the transmission spectrum. First, we consider $\omega/\omega_0 = 1.35$, where the transmission spectrum at $P_{\mathrm{in}} = 0.5$ exhibits a bistable feature for $S_{12}$, while $S_{21}$ remains close to its linear response. As a result, increasing the input power induces a pronounced nonlinear response for backward propagation, whereas the forward transmission remains only weakly affected.

At $\omega/\omega_0 = 1.4$, both propagation directions display a strong nonlinear response. However, the transition occurs at lower input powers for $S_{21}$. As the input power is increased further, the transmission associated with $S_{12}$ undergoes a more substantial increase and eventually exceeds the forward transmission.

For $\omega/\omega_0 = 1.6$, which lies close to a resonance of the linear device, both nonlinear transmission curves remain below the corresponding linear response over the investigated power range. In this regime, the nonlinear phase shifts effectively detune the resonance and thereby suppress transmission.

Finally, at $\omega/\omega_0 = 1.8$, sufficiently far from resonance, the nonlinear effects are weak and the linear and nonlinear responses nearly coincide even at the highest input powers considered.

\begin{figure}[htbp]
\centering
\includegraphics[width=1\linewidth]{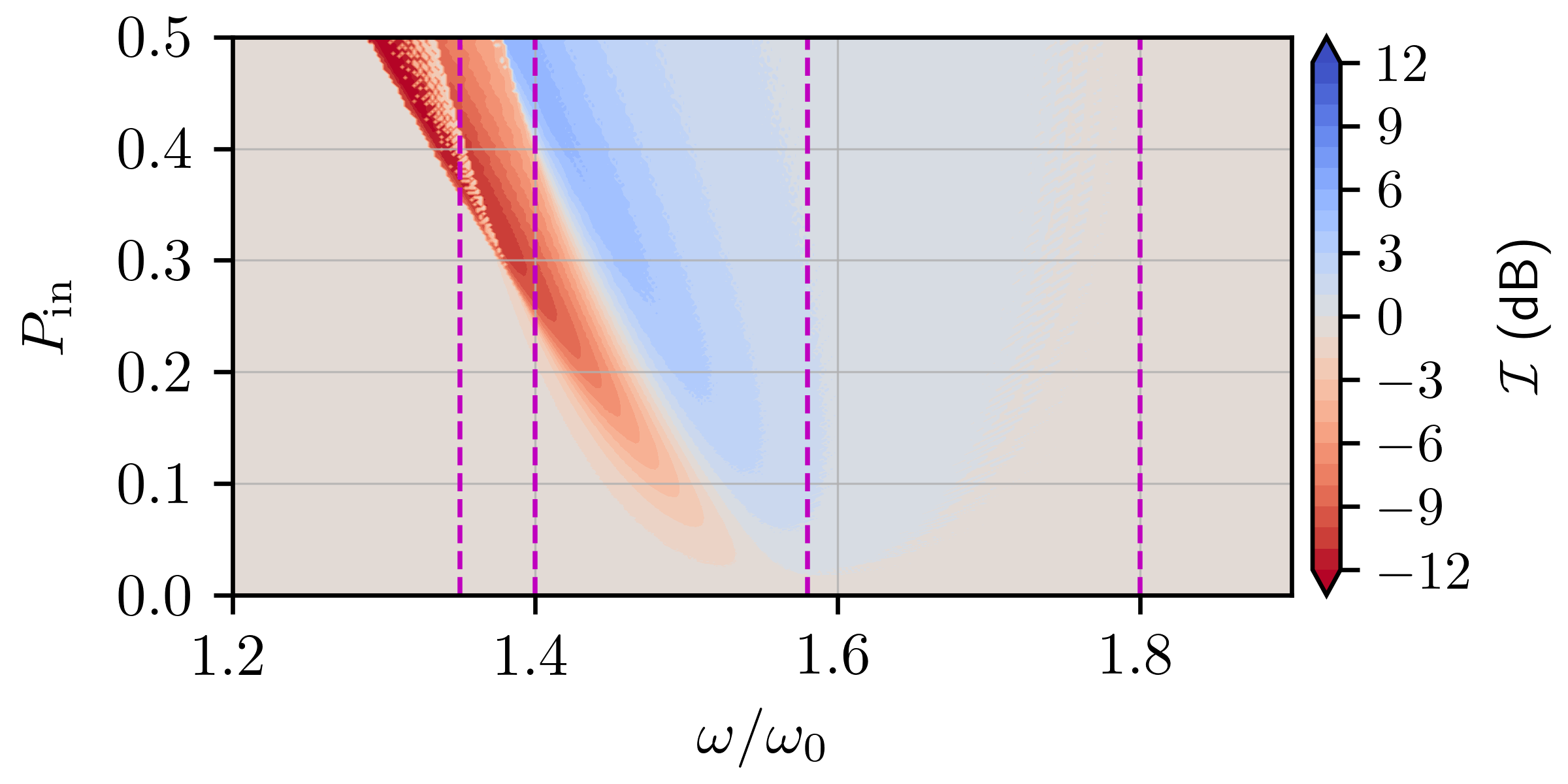}
\caption{Simulated isolation as a function of frequency and input power. The isolation $\mathcal{I}$ reveals distinct regions of strong transmission asymmetry, reaching its largest values at elevated input powers near nonlinear resonances. For reference, the four frequencies analyzed in Fig.~\ref{fig:R4}.(b-e) are indicated by dashed magenta lines. The transmission spectrum shown in Fig.~\ref{fig:R4}. (a) corresponds to the horizontal cross section at $P_{\mathrm{in}} = 0.5$.}
\label{fig:R5}
\end{figure}

As an alternative figure of merit for the device performance, we consider the isolation
\begin{equation}
    \mathcal{I} = \max \left[ 20 \log_{10} \left( \frac{|S_{21}|}{|S_{12}|} \right) \right],
\end{equation}
where the maximum is taken over all solutions of the nonlinear boundary value problem. This quantity characterizes the highest achievable directionality of the device. As illustrated in Fig.~\ref{fig:R5}, the isolation provides a compact measure of the transmission asymmetry as a function of both frequency and input power. 

Up to this point, we have considered only two-port devices composed of a single line of cascaded segments with impedance mismatches and nonlinear transmission-line sections. This naturally raises the question of whether the same formalism can be extended to multi-port geometries and, if so, what types of boundary value problems arise in physically relevant configurations.

\begin{figure}[htbp]
\centering
\includegraphics[width=1\linewidth]{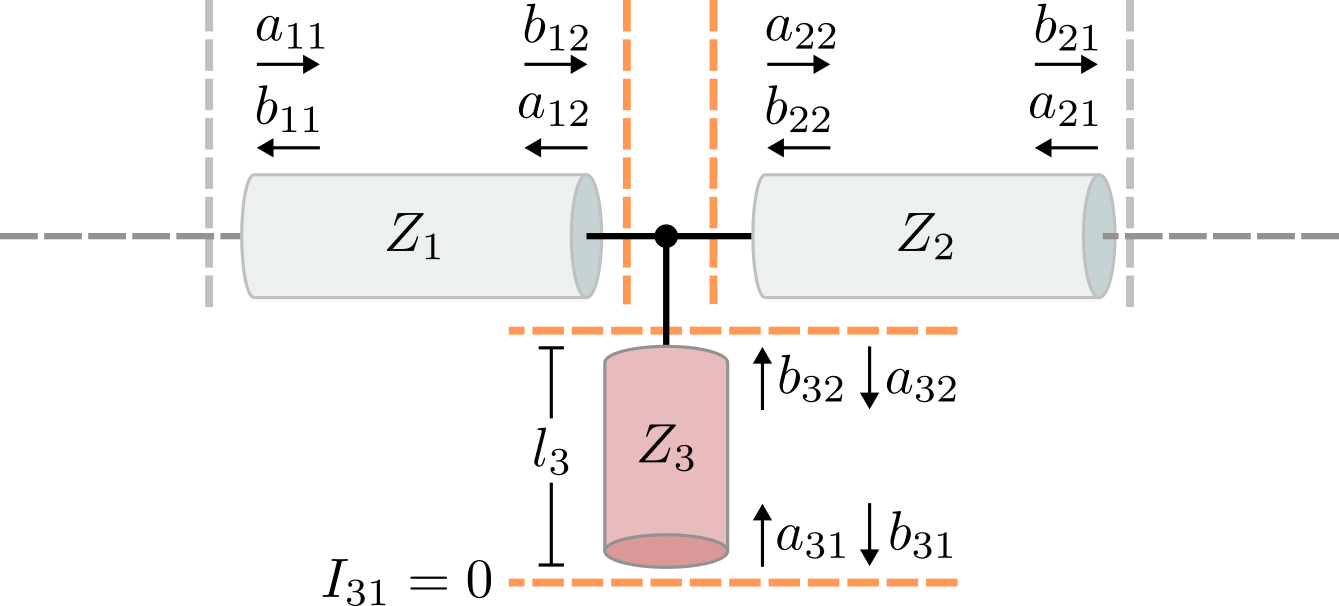}
\caption{Schematic layout of the stub geometry. The device consists of three transmission-line sections with characteristic impedances $Z_1$, $Z_2$, and $Z_3$ connected at a common node. Incoming and outgoing waves at port $m$ of section $n$ are denoted by $a_{nm}$ and $b_{nm}$, respectively. The external ports correspond to $m=1$, while the ports connected to the central node correspond to $m=2$. The sections with impedances $Z_1$ and $Z_2$ couple the device to the external environment, whereas the section with impedance $Z_3$ forms an open-circuit terminated stub.}
\label{fig:R6}
\end{figure}

As a first example, we investigate the open-ended stub geometry shown in Fig.~\ref{fig:R6}, consisting of a transmission line with a side-coupled branch. In contrast to the previously studied resonator structures, the nonlinear section is not located directly in the transmission path between the input and output ports. Instead, the stub acts as an intermediate nonlinear channel that is side-coupled to the main transmission line, resembling the role of a gate in a transistor-like architecture. This geometry therefore illustrates how nonlinear side branches can influence signal transport without being part of the direct propagation path.

The stub impedance is chosen to be one order of magnitude larger than that of the main transmission line, namely $Z_3/Z_r = 10$ and $Z_{1,2}/Z_r = 1$. Consequently, we assume that only the stub exhibits a nonlinear kinetic inductance, since the resulting nonlinear effects are expected to be most pronounced in this section.

The lower end of the stub is assumed to be open-circuit terminated. Consequently, no current can flow through the termination, yielding the boundary condition $I_{31}= (a_{31}-b_{31}) / Z_3=0$, or equivalently $a_{31}=b_{31}$. The voltage at the termination, $V_{31}=a_{31}+b_{31}$, is not fixed and instead follows self-consistently from the solution. Owing to the open-circuit boundary condition, the stub acts as a quarter-wave resonator and gives rise to transmission anti-resonances whenever its length satisfies ${\beta l_3=(2 n+1)\frac{\pi}{2}}$.

\begin{figure}[htbp]
    \centering
    \includegraphics[width=1\linewidth]{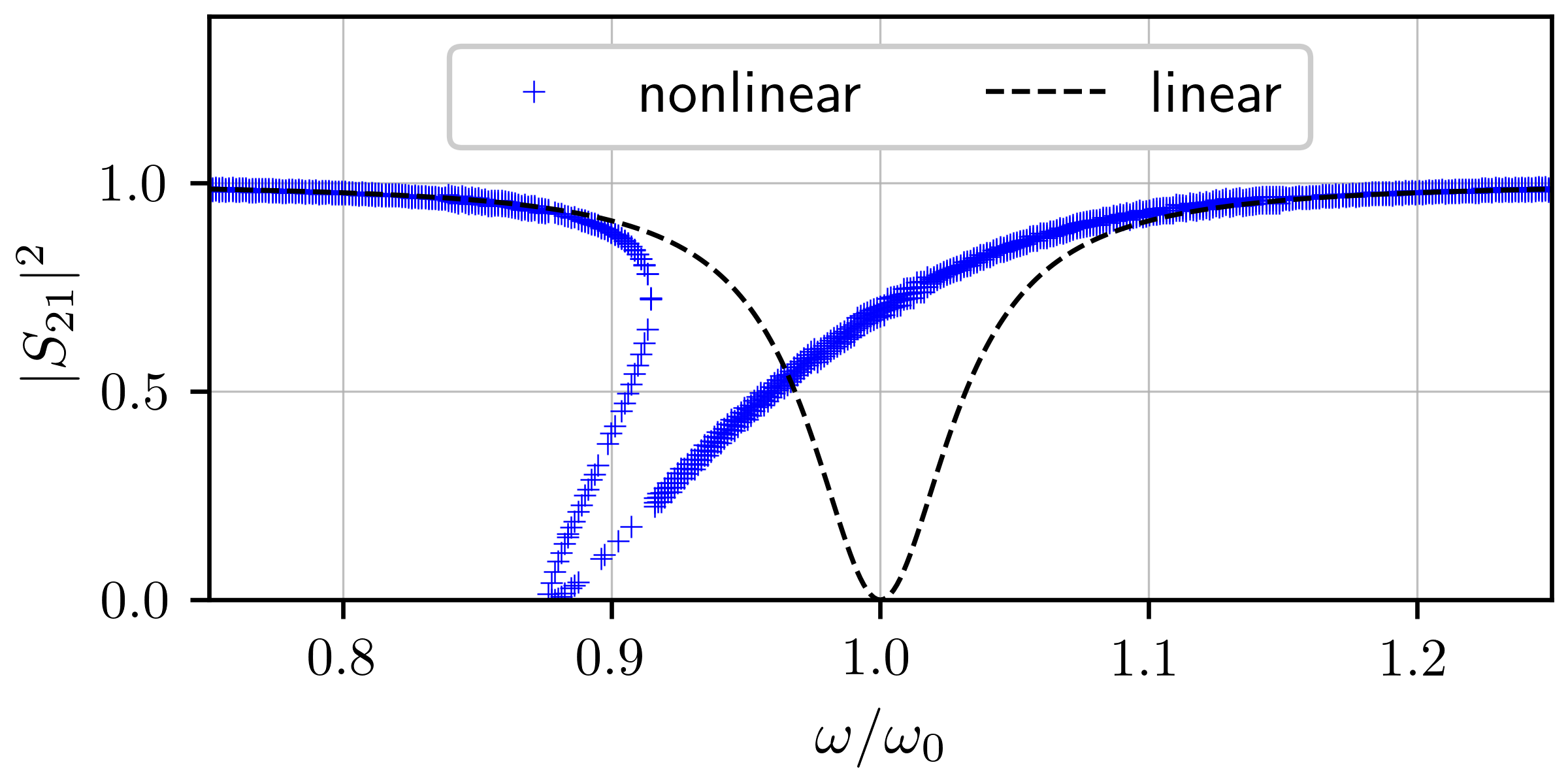}
    \caption{Transmission spectrum of the stub geometry. The characteristic impedances are chosen as $Z_{1,2}^{(0)}/Z_r = 1$ and $Z_3^{(0)}/Z_r = 10$. In the nonlinear configuration, $\xi_{1,2}=0$ and $\xi_3=0.001$, such that only the stub section exhibits a nonlinear inductance. The stub length is $\beta_3 (\omega_0) l_3=\pi/2$ (and $\beta_{1,2}(\omega_0) l_{1,2}=2\pi$). The transmission spectrum exhibits an anti-resonance at $\omega/\omega_0=1$ in the linear case (dashed black). In the nonlinear regime (blue), the anti-resonance is shifted toward lower frequencies.}
    \label{fig:R7}
\end{figure}

In the linear regime, we indeed observe a a series of transmission anti-resonances, see Fig.~\ref{fig:R7}. These arise from destructive interference between the wave propagating directly through the transmission line and the wave that enters the stub, is reflected at its open end, and subsequently re-enters the main transmission path. As the impedance ratio $Z_3/Z_{1,2}$ increases, the coupling between the stub and the transmission line becomes more selective, resulting in progressively sharper anti-resonance features.

In the nonlinear regime, the power-dependent kinetic inductance increases the effective length of the stub and shifts the anti-resonance toward lower frequencies, as shown in Fig.~\ref{fig:R7}. The resulting line shape becomes increasingly asymmetric and may develop multiple coexisting solutions, closely resembling the response of a Duffing oscillator. This similarity originates from the amplitude-dependent resonance frequency of the quarter-wave stub mode induced by the nonlinear kinetic inductance. The coexistence of multiple solution branches suggests the possibility of hysteretic transmission behavior under frequency or power sweeps, analogous to the well-known hysteresis observed in driven Duffing resonators.

The stub geometry demonstrates that the present approach can be generalized to multi-port transmission-line networks. However, the complexity of the associated boundary value problem increases rapidly with network connectivity. While the open-ended stub requires only a single additional constraint, more general multi-port devices may support multiple interacting resonant modes and coexisting nonlinear solution branches. This is particularly relevant for networks containing closed loops, where waves can propagate along different paths and repeatedly interact with nonlinear elements. Such systems are expected to exhibit considerably richer dynamical behavior and pose substantially greater numerical challenges than the serial geometries considered previously. Extending the present framework to more complex network topologies, particularly those containing closed propagation loops, therefore remains an interesting direction for future work.

%% file: sections/conclusion.tex
\section{Conclusion\label{sec:conclusion}}
We have presented a method for the analysis of nonlinear high-kinetic-inductance (HKI) transmission lines under physically relevant boundary conditions. By formulating the problem as a boundary value problem, the approach enables the treatment of strongly nonlinear transmission structures beyond perturbative descriptions based on slowly varying amplitudes and naturally incorporates resonant structures, impedance discontinuities, and strong spatial field variations.

Using this framework, we investigated several nonlinear transmission-line geometries. For resonator structures, the nonlinear kinetic inductance leads to amplitude-dependent resonance shifts and bistable transmission solutions. Introducing structural asymmetry through impedance-mismatched resonator barriers leads to direction-dependent field enhancement and unequal nonlinear phase accumulation for opposite propagation directions. As a consequence, significant transmission asymmetries and nonlinear isolation can be achieved without magnetic bias fields or active modulation, with transmission contrasts of up to $93\%$ for experimentally realistic device parameters. Furthermore, the dependence of the transmission characteristics on frequency and input power was analyzed through simulations of transmission, power-response, and isolation measurements.

The framework was also extended to a three-port stub geometry, demonstrating that nonlinear interference effects in multi-port networks can be treated within the same formalism. In this case, the side-coupled quarter-wave resonator acts as an intermediate nonlinear channel, giving rise to transmission anti-resonances through destructive interference. The nonlinear kinetic inductance shifts these anti-resonances and produces Duffing-like distortions of the spectral response, including the emergence of multiple solution branches and the possibility of hysteretic behavior.

The presented results establish a versatile framework for studying strongly nonlinear wave propagation in superconducting microwave structures and demonstrate how distributed kinetic-inductance nonlinearities can be exploited to engineer direction-dependent microwave transport while maintaining a compact device footprint and fabrication simplicity. Such devices are promising candidates for the integration of tunable microwave functionality in superconducting quantum circuits. Future work may explore optimized device geometries, stability analyses of the coexisting solution branches, and the extension of the present framework to more complex multi-port networks containing multiple resonators, junctions, and closed propagation loops.

%% file: sections/acknowledgments.tex
\begin{acknowledgments}
The authors thank I. M. Pop for valuable discussions and helpful comments on the manuscript.
This work was supported by the
Center for Integrated Quantum Science and Technology (IQST) and the Carl Zeiss Foundation through the Carl Zeiss Foundation Center QPhoton. N.K. acknowledges the support of the EU project OpenSuperQPlus.
\end{acknowledgments}

%% file: sections/appendix_a.tex
\section{Angular momentum}
The constant of motion $L_\theta$ in current and voltage representation is
\begin{equation}
    L_\theta = \omega C^{(0)} \left( I_s V_s + I_c V_c \right) \label{eq:A1}
\end{equation}
Furthermore, the voltage and current waves of a single frequency are related to their complex components by $V(x,t)=\mathrm{Re}[V(x)e^{i \omega t} ]$ and $I(x,t)=\mathrm{Re}[I(x)e^{i \omega t} ]$ respectively \cite{pozar}.
Comparing this relation with the ansatz in Eq.~\eqref{eq4}, we can identify the wave components as
\begin{subequations}\label{eq:A2}
\begin{align}
V_s &=-\mathrm{Im}\left[ V(x) \right] \label{eq:A2a}\\
V_c &=\mathrm{Re}\left[ V(x) \right] \label{eq:A2b}\\
I_s &=-\mathrm{Im}\left[ I(x) \right] \label{eq:A2c}\\
I_c &=\mathrm{Re} \left[ I(x) \right] \label{eq:A2d} 
\end{align}
\end{subequations}
Inserting Eq.~\eqref{eq:A2} into Eq.~\eqref{eq:A1} we obtain
\begin{equation}
    L_\theta = \omega C^{(0)}  \mathrm{Re} \left[ I(x) V^\ast(x) \right].\label{eq:A3}
 \end{equation}
Thus, the constant of motion that we found from the Euler-Lagrange formalism is proportional to the average power flow $P(x) = \mathrm{Re} \left[ I(x) V^\ast(x) \right]/2$.

%% file: bib.bib
@article{devoret2013,
author = {M. H. Devoret  and R. J. Schoelkopf },
title = {Superconducting Circuits for Quantum Information: An Outlook},
journal = {Science},
volume = {339},
number = {6124},
pages = {1169-1174},
year = {2013},
doi = {10.1126/science.1231930},
URL = {https://www.science.org/doi/abs/10.1126/science.1231930}
}

@article{blais2020,
       author = {{Blais}, Alexandre and {Girvin}, Steven M. and {Oliver}, William D.},
        title = "{Quantum information processing and quantum optics with circuit quantum electrodynamics}",
      journal = {Nature Physics},
         year = 2020,
        month = mar,
       volume = {16},
       number = {3},
        pages = {247-256},
          doi = {10.1038/s41567-020-0806-z}
}

@article{gu2017,
title = {Microwave photonics with superconducting quantum circuits},
journal = {Physics Reports},
volume = {718-719},
pages = {1-102},
year = {2017},
issn = {0370-1573},
doi = {https://doi.org/10.1016/j.physrep.2017.10.002},
url = {https://www.sciencedirect.com/science/article/pii/S0370157317303290},
author = {Xiu Gu and Anton Frisk Kockum and Adam Miranowicz and Yu-xi Liu and Franco Nori}
}

@article{eom2012,
author = {{Ho Eom}, Byeong and {Day}, Peter K. and {LeDuc}, Henry G. and {Zmuidzinas}, Jonas},
        title = "{A wideband, low-noise superconducting amplifier with high dynamic range}",
      journal = {Nature Physics},
         year = 2012,
        month = aug,
       volume = {8},
       number = {8},
        pages = {623-627},
          doi = {10.1038/nphys2356}
}

@article{gao2017,
    author = {Chaudhuri, S. and Li, D. and Irwin, K. D. and Bockstiegel, C. and Hubmayr, J. and Ullom, J. N. and Vissers, M. R. and Gao, J.},
    title = {Broadband parametric amplifiers based on nonlinear kinetic inductance artificial transmission lines},
    journal = {Applied Physics Letters},
    volume = {110},
    number = {15},
    pages = {152601},
    year = {2017},
    month = {04},
    issn = {0003-6951},
    doi = {10.1063/1.4980102},
    url = {https://doi.org/10.1063/1.4980102}
}

@article{malnou2021,
  title = {Three-Wave Mixing Kinetic Inductance Traveling-Wave Amplifier with Near-Quantum-Limited Noise Performance},
  author = {Malnou, M. and Vissers, M.R. and Wheeler, J.D. and Aumentado, J. and Hubmayr, J. and Ullom, J.N. and Gao, J.},
  journal = {PRX Quantum},
  volume = {2},
  issue = {1},
  pages = {010302},
  numpages = {17},
  year = {2021},
  month = {Jan},
  publisher = {American Physical Society},
  doi = {10.1103/PRXQuantum.2.010302},
  url = {https://link.aps.org/doi/10.1103/PRXQuantum.2.010302}
}

@article{makhlin2001,
 title = {Quantum-state engineering with Josephson-junction devices},
  author = {Makhlin, Yuriy and Sch\"on, Gerd and Shnirman, Alexander},
  journal = {Rev. Mod. Phys.},
  volume = {73},
  issue = {2},
  pages = {357--400},
  numpages = {0},
  year = {2001},
  month = {May},
  publisher = {American Physical Society},
  doi = {10.1103/RevModPhys.73.357},
  url = {https://link.aps.org/doi/10.1103/RevModPhys.73.357}
}

@article{annunziata2010,
doi = {10.1088/0957-4484/21/44/445202},
url = {https://doi.org/10.1088/0957-4484/21/44/445202},
year = {2010},
month = {oct},
publisher = {},
volume = {21},
number = {44},
pages = {445202},
author = {Annunziata, Anthony J and Santavicca, Daniel F and Frunzio, Luigi and Catelani, Gianluigi and Rooks, Michael J and Frydman, Aviad and Prober, Daniel E},
title = {Tunable superconducting nanoinductors},
journal = {Nanotechnology},
}

@article{vissers2016,
    author = {Vissers, M. R. and Erickson, R. P. and Ku, H.-S. and Vale, Leila and Wu, Xian and Hilton, G. C. and Pappas, D. P.},
    title = {Low-noise kinetic inductance traveling-wave amplifier using three-wave mixing},
    journal = {Applied Physics Letters},
    volume = {108},
    number = {1},
    pages = {012601},
    year = {2016},
    month = {01},
    issn = {0003-6951},
    doi = {10.1063/1.4937922},
    url = {https://doi.org/10.1063/1.4937922}
}

@article{erickson2017,
    title = {Theory of multiwave mixing within the superconducting kinetic-inductance traveling-wave amplifier},
    author = {Erickson, R. P. and Pappas, D. P.},
    journal = {Phys. Rev. B},
    volume = {95},
    issue = {10},
    pages = {104506},
    numpages = {27},
    year = {2017},
    month = {Mar},
    publisher = {American Physical Society},
    doi = {10.1103/PhysRevB.95.104506},
    url = {https://link.aps.org/doi/10.1103/PhysRevB.95.104506}
}

@article{kern2023,
  title = {Reflection-enhanced gain in traveling-wave parametric amplifiers},
  author = {Kern, S. and Neilinger, P. and Il'ichev, E. and Sultanov, A. and Schmelz, M. and Linzen, S. and Kunert, J. and Oelsner, G. and Stolz, R. and Danilov, A. and Mahashabde, S. and Jayaraman, A. and Antonov, V. and Kubatkin, S. and Grajcar, M.},
  journal = {Phys. Rev. B},
  volume = {107},
  issue = {17},
  pages = {174520},
  numpages = {9},
  year = {2023},
  month = {May},
  publisher = {American Physical Society},
  doi = {10.1103/PhysRevB.107.174520},
  url = {https://link.aps.org/doi/10.1103/PhysRevB.107.174520}
}

@article{bernier2017,
author = {{Bernier}, N.~R. and {T{\'o}th}, L.~D. and {Koottandavida}, A. and {Ioannou}, M.~A. and {Malz}, D. and {Nunnenkamp}, A. and {Feofanov}, A.~K. and {Kippenberg}, T.~J.},
        title = "{Nonreciprocal reconfigurable microwave optomechanical circuit}",
      journal = {Nature Communications},
         year = 2017,
        month = sep,
       volume = {8},
          eid = {604},
        pages = {604},
          doi = {10.1038/s41467-017-00447-1}
}

@article{abdo2017,
  title = {Gyrator Operation Using Josephson Mixers},
  author = {Abdo, Baleegh and Brink, Markus and Chow, Jerry M.},
  journal = {Phys. Rev. Appl.},
  volume = {8},
  issue = {3},
  pages = {034009},
  numpages = {10},
  year = {2017},
  month = {Sep},
  publisher = {American Physical Society},
  doi = {10.1103/PhysRevApplied.8.034009},
  url = {https://link.aps.org/doi/10.1103/PhysRevApplied.8.034009}
}

@article{lecocq2017,
  title = {Nonreciprocal Microwave Signal Processing with a Field-Programmable Josephson Amplifier},
  author = {Lecocq, F. and Ranzani, L. and Peterson, G. A. and Cicak, K. and Simmonds, R. W. and Teufel, J. D. and Aumentado, J.},
  journal = {Phys. Rev. Appl.},
  volume = {7},
  issue = {2},
  pages = {024028},
  numpages = {17},
  year = {2017},
  month = {Feb},
  publisher = {American Physical Society},
  doi = {10.1103/PhysRevApplied.7.024028},
  url = {https://link.aps.org/doi/10.1103/PhysRevApplied.7.024028}
}

@ARTICLE{oates1993,
  author={Oates, J.H. and Shin, R.T. and Oates, D.E. and Tsuk, M.J. and Nguyen, P.P.},
  journal={IEEE Transactions on Applied Superconductivity}, 
  title={A nonlinear transmission line model for superconducting stripline resonators}, 
  year={1993},
  volume={3},
  number={1},
  pages={17-22},
  doi={10.1109/77.233413}}

@article{goldstein2022,
    doi = {10.1088/1367-2630/ac45cc},
    url = {https://doi.org/10.1088/1367-2630/ac45cc},
    year = {2022},
    month = {feb},
    publisher = {IOP Publishing},
    volume = {24},
    number = {2},
    pages = {023022},
    author = {Goldstein, Samuel and Pardo, Guy and Kirsh, Naftali and Gaiser, Niklas and Padurariu, Ciprian and Kubala, Björn and Ankerhold, Joachim and Katz, Nadav},
    title = {Compact itinerant microwave photonics with superconducting high-kinetic inductance microstrips},
    journal = {New Journal of Physics}
}

@article{zmuidzinas2012,
author = "Zmuidzinas, Jonas",
   title = "Superconducting Microresonators: Physics and Applications", 
   journal= "Annual Review of Condensed Matter Physics",
   year = "2012",
   volume = "3",
   number = "Volume 3, 2012",
   pages = "169-214",
   doi = "https://doi.org/10.1146/annurev-conmatphys-020911-125022",
   url = "https://www.annualreviews.org/content/journals/10.1146/annurev-conmatphys-020911-125022",
   publisher = "Annual Reviews",
   issn = "1947-5462",
   type = "Journal Article"
  }

@book{pozar,
	author = {Pozar, David M.},
	title = {Microwave engineering},
	publisher = {John Wiley \& Sons, Inc},
	year = {2012},
	address = {New York},
	edition = {4th ed.}
}

@article{yang2024,
author = {Yang, Ming and He, XiaoLiang and Gao, WanPeng and Chen, JunFeng and Wu, Yu and Wang, XiaoNi and Mu, Gang and Peng, Wei and Lin, ZhiRong},
    title = {Kinetic inductance compact resonator with NbTiN micronwires},
    journal = {AIP Advances},
    volume = {14},
    number = {8},
    pages = {085027},
    year = {2024},
    month = {08},
    issn = {2158-3226},
    doi = {10.1063/5.0220296},
    url = {https://doi.org/10.1063/5.0220296}
}

@article{frasca2023,
  title = {NbN films with high kinetic inductance for high-quality compact superconducting resonators},
  author = {Frasca, S. and Arabadzhiev, I.N. and de Puechredon, S.Y. Bros and Oppliger, F. and Jouanny, V. and Musio, R. and Scigliuzzo, M. and Minganti, F. and Scarlino, P. and Charbon, E.},
  journal = {Phys. Rev. Appl.},
  volume = {20},
  issue = {4},
  pages = {044021},
  numpages = {13},
  year = {2023},
  month = {Oct},
  publisher = {American Physical Society},
  doi = {10.1103/PhysRevApplied.20.044021},
  url = {https://link.aps.org/doi/10.1103/PhysRevApplied.20.044021}
}

@phdthesis{gao2008,
    author = {{Gao}, Jiansong},
    title = {{The Physics of {Superconducting} {Microwave} {Resonators}}},
    school = {California Institute of Technology},
    year = 2008,
    month = jan
}

@article{yurke1989,
  title = {Observation of parametric amplification and deamplification in a Josephson parametric amplifier},
  author = {Yurke, B. and Corruccini, L. R. and Kaminsky, P. G. and Rupp, L. W. and Smith, A. D. and Silver, A. H. and Simon, R. W. and Whittaker, E. A.},
  journal = {Phys. Rev. A},
  volume = {39},
  issue = {5},
  pages = {2519--2533},
  numpages = {0},
  year = {1989},
  month = {Mar},
  publisher = {American Physical Society},
  doi = {10.1103/PhysRevA.39.2519},
  url = {https://link.aps.org/doi/10.1103/PhysRevA.39.2519}
}

@article{lehnert2007,
    author = {Castellanos-Beltran, M. A. and Lehnert, K. W.},
    title = {Widely tunable parametric amplifier based on a superconducting quantum interference device array resonator},
    journal = {Applied Physics Letters},
    volume = {91},
    number = {8},
    pages = {083509},
    year = {2007},
    month = {08},
    issn = {0003-6951},
    doi = {10.1063/1.2773988},
    url = {https://doi.org/10.1063/1.2773988}}

@article{macklin2015,
    author = {C. Macklin  and K. O’Brien  and D. Hover  and M. E. Schwartz  and V. Bolkhovsky  and X. Zhang  and W. D. Oliver  and I. Siddiqi },
    title = {A near–quantum-limited Josephson traveling-wave parametric amplifier},
    journal = {Science},
    volume = {350},
    number = {6258},
    pages = {307-310},
    year = {2015},
    doi = {10.1126/science.aaa8525},
    URL = {https://www.science.org/doi/abs/10.1126/science.aaa8525}
}

@article{goldstein2020,
    author = {Goldstein, Samuel and Kirsh, Naftali and Svetitsky, Elisha and Zamir, Yuval and Hachmo, Ori and de Oliveira, Clovis Eduardo Mazzotti and Katz, Nadav},
    title = {Four wave-mixing in a microstrip kinetic inductance travelling wave parametric amplifier},
    journal = {Applied Physics Letters},
    volume = {116},
    number = {15},
    pages = {152602},
    year = {2020},
    month = {04},
    issn = {0003-6951},
    doi = {10.1063/5.0004236},
    url = {https://doi.org/10.1063/5.0004236}
}

@book{nayfeh2024,
  title={Nonlinear oscillations},
  author={Nayfeh, Ali H and Mook, Dean T},
  year={2024},
  publisher={John Wiley \& Sons}
}

@book{guckenheimer1983,
author="Guckenheimer, John
and Holmes, Philip",
Title="Nonlinear Oscillations, Dynamical Systems, and Bifurcations of Vector Fields",
year="1983",
publisher="Springer New York"}

@book{keller2018,
  title={Numerical Methods for Two-Point Boundary-Value Problems},
  author={Keller, H.B.},
  isbn={9780486828343},
  lccn={2018019764},
  series={Dover Books on Mathematics},
  year={2018},
  publisher={Dover Publications}
}

@book{ascher1995,
author = {Ascher, Uri M. and Mattheij, Robert M. M. and Russell, Robert D.},
title = {Numerical Solution of Boundary Value Problems for Ordinary Differential Equations},
publisher = {Society for Industrial and Applied Mathematics},
year = {1995},
doi = {10.1137/1.9781611971231},
URL = {https://epubs.siam.org/doi/abs/10.1137/1.9781611971231}
}

@ARTICLE{cotrufo2021,
  author={Cotrufo, Michele and Mann, Sander A. and Moussa, Hady and Alù, Andrea},
  journal={IEEE Transactions on Microwave Theory and Techniques}, 
  title={Nonlinearity-Induced Nonreciprocity—Part I}, 
  year={2021},
  volume={69},
  number={8},
  pages={3569-3583},
  doi={10.1109/TMTT.2021.3079250}}
